\documentclass[aps,twocolumn,showpacs,preprintnumbers,amsmath,amssymb,pre]{revtex4}
\usepackage{epsf,amsmath,amssymb,verbatim,color,multirow,pifont}
\usepackage{graphicx}

\newcommand{\SCE}{Smoluchowski coagulation equation}
\newcommand{\CAN}{cluster aggregation network}

\begin{document}
\title{Cluster aggregation model for discontinuous percolation transition}
\author{Y.S.~Cho, B.~Kahng and D.~Kim}
\affiliation{Department of Physics and Astronomy, Seoul National University,
Seoul 151-747, Korea}
\date{\today}

\begin{abstract}
The evolution of the Erd\H{o}s-R\'enyi (ER) network by adding
edges is a basis model for irreversible kinetic aggregation
phenomena. Such ER processes can be described by a rate equation
for the evolution of the cluster-size distribution with the
connection kernel $K_{ij}\sim ij$, where $ij$ is the product of
the sizes of two merging clusters. Here we study that when the
giant cluster is discouraged to develop by a sub-linear kernel
$K_{ij}\sim (ij)^{\omega}$ with $0 \le \omega < 1/2$, the
percolation transition (PT) is discontinuous. Such discontinuous
PT can occur even when the ER dynamics evolves from proper initial
conditions. The obtained evolutionary properties of the simple model
sheds light on the origin of the discontinuous PT in other
non-equilibrium kinetic systems.
\end{abstract}

 \maketitle

Irreversible cluster aggregations are widespread phenomena
occurring in a diverse range of fields, including dust and colloid
formation, aerosol growth, droplet nucleation and growth, gelation
transition, etc~\cite{review}. The Smoluchowski coagulation
equation~\cite{smo1,flory,stockmayer} can successfully describe
such cluster aggregation processes. In linear polymerization,
molecules with two reactive ends can react to form long chains. In
this case, the reaction kernel is given as $K_{ij}=1$, where $i$
and $j$ are the masses of the two reactants. For the aggregation
of branched polymers, the reaction kernel has the form
$K_{ij}=(ai+b)(aj+b)$, where $a$ and $b$ are constants. When
clusters have a compact shape, the reaction kernel has the form
$K_{ij}\sim (ij)^{1-1/d}$, where $d$ is the spatial dimension.
Intensive studies have been carried out using the \SCE~with such
different kernel types~\cite{review,ziff0,ziff1,levyraz,phyrep},
and it is known that sol-gel transitions can occur at either
finite or infinite transition points. They are continuous
transitions.

\begin{figure}[t]
\includegraphics[width=1.0\linewidth]{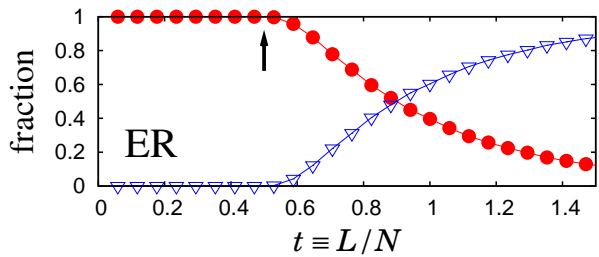}
\caption{(Color online) The fraction of each type of attached
edges, inter-cluster ($\bullet$) or intra-cluster
($\bigtriangledown$) edges for the ER model. Arrow indicates
percolation threshold at $p_c=1/2$.}\label{events}
\end{figure}

During the past decade, the evolution of complex networks has been
of much interest to the science communities in multidisciplinary
fields. To study percolation transition (PT) during network
evolution, the branching process approach~\cite{newman,havlin} and
the Potts model formalism~\cite{dslee} have been used. Such
complex network evolution can also be viewed as a cluster
aggregation phenomenon, and can be studied by the
rate-equation approach~\cite{ziff1}. For example, in the evolution
of the classical random network, called the Erd\H{o}s-R\'enyi (ER)
model, an edge is added at each time step, thereby either
connecting two separate clusters (inter-cluster edge) or
increasing the edge number in one cluster without changing cluster
numbers (intra-cluster edge). Fig.~\ref{events} shows that the
frequency of inter-cluster connections is dominant until the
percolation threshold. Thus, the cluster aggregation picture of
the ER network evolution comes in naturally. In this paper, we
extend the cluster aggregation dynamics in networks to more
general cases. Specifically, the model is as follows: In a system
composed of $N$ vertices, we perform the following tasks at each
time step.
\begin{itemize}
\item[$\bullet$] Two clusters of sizes $i$ and $j$ are chosen with
probabilities $q_i$ and $q_j$, respectively. The two clusters can
be the same. Probability $q_i$ is given as $k_i/\sum_s k_s n_s$,
where $k_i$ and $n_i$ are the weight and density of an $i$-sized
cluster, respectively. \item[$\bullet$] Two vertices are selected
randomly one each from the selected clusters. If they are not yet
connected, then they are connected by an edge. If they are already
connected, we choose another pair of vertices in the same manner
until a link can be added. Self-loop cases are excluded.
\end{itemize}
We repeat these simple steps until a given time $t\equiv L/N$,
where $L$, the number of edges added to the system, is tuned. This
model is called the cluster aggregation network model hereafter.
In this model, the two selected clusters can be the same, and
thus, the evolution can proceed even after one giant cluster
remains. The ER network corresponds to the case $k_i=i$. Here, we
show that when the weight is sub-linear, as $k_i=i^{\omega}$ with
$0 \le \omega < 1/2$, a discontinuous PT occurs at a finite
transition point. Moreover, under certain initial conditions, the
ER dynamics also exhibits a discontinuous PT. This observation is
remarkable, because a discontinuous PT has rarely been discovered
in irreversible kinetic systems, except for recent observations in
the ER~\cite{science} and other networks~\cite{ziff_prl,cho,santo,friedman,herrmann,makse}
under the so-called Achlioptas process~\cite{science}. On the
other hand, it is noteworthy that the \CAN~model evolves by
single-edge dynamics, as compared with the ER network under the
Achlioptas process, which involves a pair of edges at each time
step. Thus, this \CAN~model allows us to study the underlying
mechanism of the discontinuous PT analytically for some cases,
which is shown later.

The cluster aggregation processes in the model are described via a
rate equation for the cluster density, which takes the following
form in the thermodynamic limit:
\begin{equation}
\frac{dn_{s}(t)}{dt}=\sum_{i+j=s}\frac{k_i n_i}{c(t)}\frac{k_j n_j}
{c(t)}-2\frac{k_s n_{s}}{c(t)},\label{rate}
\end{equation}
where $c(t)=\sum_s k_s n_s(t)$. The connection kernel
$K_{ij}\equiv k_ik_j/c^2$. The first term on the right hand side
represents the aggregation of two clusters of sizes $i$ and $j$
with $i+j=s$ and the second term represents a cluster of size $s$
merging with another cluster of any size. The rate equation
differs from the \SCE~in two aspects. First, the connection kernel
is time-dependent through $c(t)$ when $\omega \ne 1$. Second, the
second term on the right hand side of Eq.~(\ref{rate}) includes
the process of merging with an infinite-size cluster. Hence,
Eq.~(\ref{rate}) with $\omega=1$ and $c=1$ describes the ER
process, while the conventional \SCE~with $\omega=1$ does not,
because only sol-sol reactions are taken into account. However,
the case including the infinite-size cluster in the \SCE~was also
considered in Ref~\cite{ziff1}, which was called the F-model.
Owing to the presence of $c(t)$, a PT occurs at a finite
transition point even when a PT does not occur in the \SCE, for
example, when $\omega=0$. Here, we study the cases $k_i=1$
($\omega=0$), $k_i=i^{\omega}$ with $0< \omega < 1$, and $k_i=i$
($\omega=1$), separately.

\begin{figure}[t]
\includegraphics[width=0.8\linewidth]{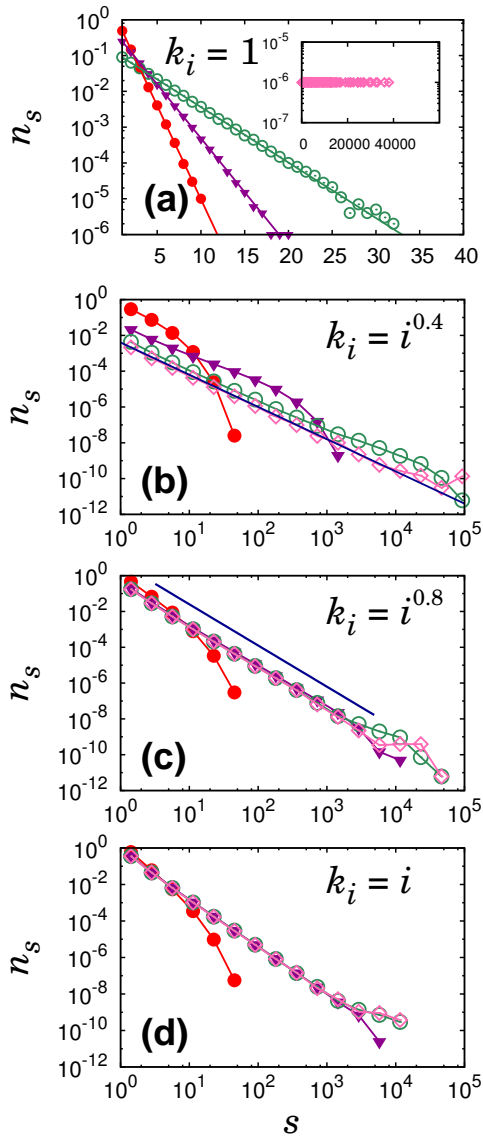}
\caption{(Color online) The cluster-size distributions of the
\CAN~models with (a) $k_{i}=1$, (b) $k_i=i^{0.4}$, (c) $k_i=i^{0.8}$,
and (d) $k_i=1$ (ER model) with mono-disperse initial condition.
(a) is drawn in semi-logarithmic scale, while the others are
in double-logarithmic scales. Data points in (a) are at
$t/t_c=0.3 (\bullet)$, $0.5
(\triangledown)$ and $0.7 (\circ)$, and in (b), (c), and (d), they
are obtained at $t/t_c=0.50~(\bullet)$, $0.95~(\triangledown)$,
$0.998$~($\circ$)), and $1.003$ ($\diamond$)). Data points in the
inset of (a) are at $\delta=1-t=10^{-4}$. In (b), there exists
a hump for data points ($\circ$). The system size $N$ is taken as
$10^6$ for (a) and $10^5$ for (b), (c), and (d).
Solid lines in (b) and (c) represent analytic formulae Eq.~(\ref{tau_omega})}
\label{cluster}
\end{figure}

{\sl The case $\omega=0$:} In this case, $c(t)=\sum_{s}n_s$
becomes the total density of the clusters, which decreases
linearly with time. The generating function of $n_{s}(t)$ is
defined as $f(z,t)=\sum_{s}n_{s}(t)z^{s}$, where $z$ is the
fugacity in the range $0 < z <1$. Then, one can obtain the
differential equation for $f(z,t)$ from Eq.~(\ref{rate}) and solve
in a closed form as $f(z,t)={(1-t)^{2}z}/(1-zt)$ for $t<1$ and 0
for $t> 1$ in the thermodynamic limit. Expanding $f(z,t)$ as a
series in $z$, we obtain
\begin{equation}
n_{s}(t)=(1-t)^{2}t^{s-1}
\end{equation}
for $t<1$. This
formula shows that the cluster size distribution decays
exponentially as $s$ becomes large. Particularly, when $\delta
\equiv 1-t$ is small, $n_s(\delta)\approx \delta^2 e^{-s/s^*}$
with $s^*\approx 1/\delta.$ The characteristic size $s^*$ diverges
as $\delta \to 0$. As shown in the inset of Fig.~\ref{cluster}(a),
$n_s(t)$ is almost flat at $\delta=10^{-4}$ for $N=10^6$,
indicating that large-size clusters are relatively abundant. The
merging of these clusters causes a sudden jump in the giant
cluster size, leading to a first-order transition.

\begin{figure}[t]
\includegraphics[width=1.0\linewidth]{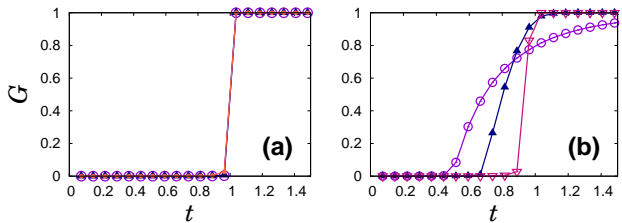}
\caption{(Color online) The density of the giant cluster size $G$
versus time $t$ showing (a) discontinuous transitions in $k_i=1$
$(\circ)$, $k_i=i^{0.2}$ $(\triangle)$, $k_i=i^{0.4}$ $(\bullet)$,
and (b) continuous transitions in the \CAN~model $k_i=i^{0.6}$
$(\triangledown)$, $k_i=i^{0.8}$ $(\triangle)$, and the ER network
($\circ$). $N=10^5$ in both (a) and (b).}\label{giant}
\end{figure}

We find the giant cluster size $G(t)$ by using the relation,
$G(t)=1-f^{\prime}(1,t)\equiv 1-\sum_s^{\prime} sn_s(t)$, where
the summation excludes an infinite-size cluster. We find that
\begin{equation}
G(t) = \left\{\begin{array}{lll}
0 & \textrm{if~} & 0 <t <1, \\
1 & \textrm{if~} & t > 1,
\end{array}\right.
\end{equation}
in the thermodynamic limit (Fig.~\ref{giant}(a)). Thus, the PT is
first-order at $t_c=1$. This result differs from what we obtain
from the \SCE, in which the transition point $t_c=\infty$.

{\sl The case $0< \omega < 1$:} For this case, while exact
solution for $n_s(t)$ is not obtained, $n_s(t_c)$ is done under
certain assumptions. To proceed, we define the generating function
$g_{\omega}(\mu,t) \equiv \sum_s s^{\omega} n_s(t)e^{\mu s}/c(t)$
($\mu <0$), and presume that $n_s(t_c)\sim s^{-\tau}$. Next, we
use the assumption made in Ref.~\cite{ziff0,levyraz} for the
\SCE~that $n_s(t)=n_s(t_c)/(1+b(t-t_c))$ near $t=t_c^+$, where $b$
is an $s$-independent constant. Then, comparing the most singular
terms in the series of the generating functions $f(e^{\mu},t_c)$
and $g_{\omega}^2({\mu},t_c)$ in $\mu$, we find that
\begin{equation}
\tau = \left\{\begin{array}{lll}
1+2\omega & \textrm{if~} & 0 < \omega < 1/2, \\
3/2+\omega & \textrm{if~} & 1/2 < \omega < 1.
\end{array}\right.\label{tau_omega}
\end{equation}
This result is confirmed numerically in Fig~\ref{cluster}. When $t <
t_c$, $n_s(t)$ follows a power-law function with an exponential
cutoff for $1/2 < \omega < 1$, but it exhibits a hump in a
large-size region for $0 < \omega < 1/2$ (Fig.~\ref{cluster}(b)).

\begin{figure}[t]
\includegraphics[width=1.0\linewidth]{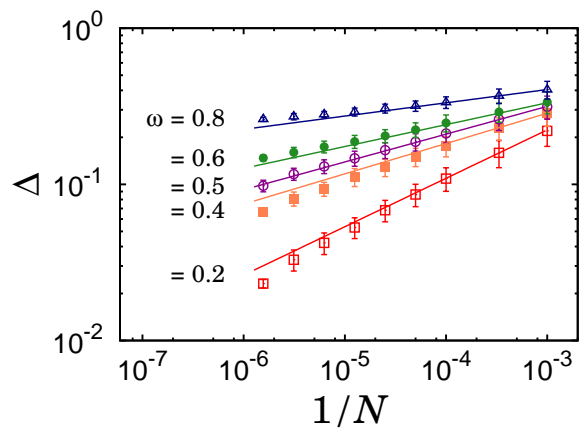}
\caption{(Color online) Test of the discontinuous or continuous PT
for the \CAN~models with various $\omega$ values. When $0< \omega < 0.5$,
$\Delta$ decays with increasing $N$; however, when $0.5 < \omega <1$,
it converges to a finite value. The straight lines are guidelines
for eyes.}\label{first}
\end{figure}

We examine $G(t)$ as a function of time for various $\omega$
cases. $G(t)$ exhibits a transition at finite $t_c$, which is
continuous for $1/2 < \omega \le 1$, discontinuous for $0 \le
\omega < 1/2$ (Fig.~\ref{giant}), and marginal for $\omega=1/2$.
The first-order transition is tested in Fig.~\ref{first} using the
scaling approach introduced in Ref.~\cite{science}. We define
$\Delta\equiv t_1-t_0$, where $t_0$ and $t_1$ are chosen as the
times at which the value of $G(t)$ reaches $1/\sqrt{N}$ and 0.8
for the first time, respectively. We find numerically that
for $0 \le \omega < 0.5$, $\Delta$ decays as $N\to \infty$,
while for $0.5 < \omega \le 1$, $\Delta$ converges to a finite value.
This result suggests that the transition is discontinuous (continuous)
for $0 \le \omega < 0.5$ ($0.5 < \omega \le 1$).

\begin{figure}[t]
\includegraphics[width=1.0\linewidth]{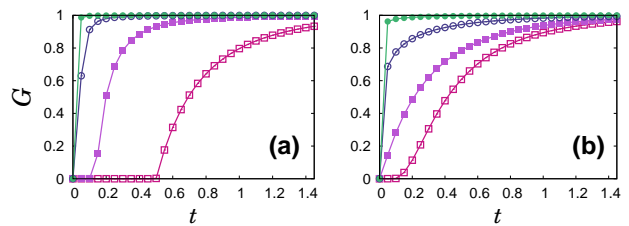}
\caption{(Color online) Plot of the giant cluster
size $G$ versus time for the ER network with different initial
conditions. In (a), the cluster-size distribution $n_s(0)$ is flat
with different cutoff values $s_m=N^{\eta}$, with $\eta=0$, 0.1,
0.2, and 0.3 from right to left. In (b), $n_s(0)$ decays according
to a power law with exponent $\tau=3$, 2.5, 2, and 1.5 from right
to left. $N=10^7$ in both (a) and (b).}\label{initial}
\end{figure}

\begin{table*}
\caption{When the number of cluster sizes at initial time obeys a
power law $n_s(0)=As^{-\tau}$ for $s=1,\dots, s_m$, listed are the
amplitude $A$, the second moment at initial time $M_2(0)$, the
critical point $t_c$, and the critical behavior of
the giant cluster size $G(t)$. Type of
PT is specified for each case. The listed $t_c$ and $G$ are the ones in the
thermodynamic limit.}
\begin{center}
\begin{tabular}{lcccccc}
\hline\hline ~~~~~~&~~~~~$\tau$~~~~~ &~~~~~A~~~~~&
~~~~~~$M_2(0)$~~~~~~
& ~~~~~$t_c$~~~~~& ~~~~~~~~$G(t > t_c)$~~~~~~~~ & ~~~{\rm type of PT}~~~
\\
\hline (i)& $0 \le \tau < 2$ & $\frac{2-\tau}{s_m^{2-\tau}}$ &
$\frac{(2-\tau)s_m}{3-\tau}$ & 0 & 1 & discontinuous
\\
(ii)& $\tau=2$ & $\frac{1}{\ln s_m}$ & $\frac{s_m}{\ln s_m}$ & 0 & 1 & discontinuous \\
(iii) & $2<\tau <3$ & $\frac{1}{\zeta(\tau-1)}$ &
$\frac{s_m^{3-\tau}}{\zeta(\tau-1)(3-\tau)}$ & 0 & $\propto
t^{\frac{\tau-2}{3-\tau}}$
& continuous \\
(iv)& $\tau=3$ & $\frac{1}{\zeta(\tau-1)}$ & $\frac{\ln
s_m}{\zeta(2)}$ & 0 & $\propto \frac{1}{2t}e^{-\zeta(2)/2t}$ & continuous \\
(v)& $3< \tau <4$ & $\frac{1}{\zeta(\tau-1)}$ &
$\frac{\zeta(\tau-2)}{\zeta(\tau-1)}$ &
$\frac{\zeta(\tau-1)}{2\zeta(\tau-2)}$ &
$\propto (t-t_c)^{\frac{1}{\tau-3}}$ & continuous \\
(vi)& $\tau=4$ & $\frac{1}{\zeta(\tau-1)}$ &
$\frac{\zeta(2)}{\zeta(3)}$  &  $\frac{\zeta(3)}{2\zeta(2)}$ &
$\propto \frac{t-t_c}{\ln(t-t_c)}$ & continuous \\
(vii) & $\tau > 4$ & $\frac{1}{\zeta(\tau-1)}$ &
$\frac{\zeta(\tau-2)}{\zeta(\tau-1)}$ &
$\frac{\zeta(\tau-1)}{2\zeta(\tau-2)}$ & $\propto t-t_c$ &
continuous
\\
\hline
\end{tabular}
\end{center}
\label{table1}
\end{table*}

{\sl The case $\omega=1$:} This case is exactly solvable as the
case of the \SCE~\cite{ziff1}. We consider an arbitrary initial
condition of $n_s(0)$. In this case, $c(t)=\sum_s sn_s(t)$ is
conserved as $c(t)=1$, but the first moment
$M_1(t)=\sum_s^{\prime} s n_s(t)$, with the sum excluding the
largest cluster, is not. The generating function
$g_1(\mu,t)=\sum_s s n_s(t) \exp(\mu s)$ satisfies the relation
\begin{equation}
\dot g_1=2(g_1-1)g_1^{\prime},
\end{equation}
where the dot (prime) is the derivative with respect to time $t$
($\mu$). Then, $g_1$ is the solution of
\begin{equation}
g_1(\mu,t)=1-H(-\mu-2t(g_1(\mu,t)-1)),\label{self-consistent}
\end{equation}
where $H(\mu)\equiv 1-g_1(-\mu,0)$ is fixed by
the initial conditions of $n_s(0)$. The giant cluster size $G(t)$,
defined as $G=1-g_1(0^-,t)$, can be solved by the self-consistent
equation $G=H(2tG)$. The obtained $G(t)$ has the form near $t_c$
as
\begin{equation}
G(t)=\frac{2M_2^2(0)}{M_3(0)}\left(2M_2(0)t-1\right),\label{giant_er}
\end{equation}
where $M_n(0)=\sum_s^{\prime} s^n n_s(0)$ is the initial $n$-th
moment. Also, one finds that the second moment $M_2(t)\equiv
\sum_s^{\prime} s^2 n_s(t)$, obtained from $g_1^{\prime}(0^-,t)$, behaves
as
\begin{equation}
M_2(t)=\frac{M_2(0)}{|1-2M_2(0)t|}\label{second_er}
\end{equation}
for $t< t_c=1/(2M_2(0))$, and for $t> t_c$ as $t \to t_c^+$.

These solutions for arbitrary initial conditions are used to study
the first-order transition in the ER network below. For the
\CAN~model, the initial condition is $n_s(0)=\delta_{s,1}$. Then,
$M_2(0)=M_3(0)=1$, and consequently, $t_c=1/2$, which is the
well-known ER value. The giant cluster size exhibits a continuous
transition at $t_c$ with $n_s(t_c)\sim s^{-5/2}$.

It is often the case that starting from $n_s(0)=\delta_{s,1}$,
the cluster-size distribution $n_s(t)$ exhibits a power-law behavior
(or with hump) in $s$ just before or at the transition point,
even when the dynamics is different from the ER (see Fig.~\ref{cluster}).
To see how such $n_s$ evolves under the ER dynamics from then on,
we consider here two particular cases in which $M_2(0)$ and $M_3(0)$
depend on $N$. First, we assume that $n_s(0)$ follows a flat distribution,
$n_s(0)=n_0$, in the range $0 < s < s_{m}$,
where $s_m$, the size of the largest cluster at $t=0$, depends on
$N$ as $s_{m}=N^{\eta}$. Then, $n_0=2N^{-2\eta}$, $M_2(0)\propto
N^{\eta}$, and $M_3(0)\propto N^{2\eta}$. Then, a PT takes place
at $t_c(N)=1/2M_2(0)\propto N^{-\eta}$, and $G(t)\sim
r(2M_2(0)t-1)$ for $t > t_c(N)$ from Eq.~(\ref{giant_er}), where
$r$ turns out to be in $\mathcal{O}(1)$. Thus, if time $t$ is
scaled as $t^{\prime}=tM_2(0)$, then one can show that $G(t^{\prime})$
is the solution of $G=\tilde{H}(3t^{\prime}G)$, where
$\tilde{H}(x)=2\sum_{n=1}^{\infty}(-1)^{n+1}x^n/(n!(n+2))$ is a
regular function qualitatively similar to $H(x)=1-e^{-x}$ of the
standard ER problem. Hence, $G(t^{\prime})$ has a mean field
behavior similar to the original ER case. This scaling behavior
implies that while $\delta G(t)\equiv G(t_1)-G(t_0)$ increases by
$\mathcal{O}(1)$, $\Delta\equiv t_1-t_0$ does so by $\sim
\mathcal{O}(N^{-\eta})$. Thus, we have a first-order transition as
$N \to \infty$ (Fig.~\ref{initial}(a)).

Second, we suppose that the initial condition is given as
$n_s(0)=A s^{-\tau}$ in the range $0 < s < s_m$, where $A$ is the
normalization constant determined by the condition $\sum_s
sn_s=1$. $A$ is given in Table I for various ranges of $\tau$, together with
the initial second moment for arbitrary $s_m$, the critical point,
and the giant cluster size $G(t)$, obtained from Eq.~(\ref{self-consistent})

In short, the PT occurs at $t=0$ for $\tau \le 3$, and at finite
$t_c$ for $\tau > 3$. This behavior is related to the divergence of
the second moment $M_2(0)$ since time is scaled in the form
$t^{\prime}=2tM_2(0)$. The transition is
discontinuous when $\tau < 2$, but continuous when $2 < \tau < 4$.
This difference originates from the fact that the ratio
$M_2^2(0)/M_3(0)$ is finite for the former, while it vanishes for
the latter. For $\tau > 4$, both $M_2(0)$ and $M_3(0)$ are finite,
resulting in the classical percolation behavior at a finite $t_c$.

In summary, we have introduced a cluster aggregation network
model, in which discontinuous percolation transitions occur when
the connection kernel is sub-linear as $K_{ij}\sim (ij)^{\omega}$
with $0 \le \omega < 1/2$. Even for the ER network, a
discontinuous PT can also be obtained by using initial conditions
where $M_2(0)$ diverges and $M_2^2(0)/M_3(0)$ remains finite.
The simple model manifests explicitly the role of the abundance of
large-size clusters just before a transition point as a mechanism of the
discontinuous PT~\cite{friedman}. We expect that the \CAN~model can
be used to study underlying dynamics of the explosive percolation
transition of random ER network under the Achlioptas process~\cite{science}.\\

This work was supported by a KOSEF grant Acceleration Research (CNRC)
(Grant No.R17-2007-073-01001-0) and by the NAP of KRCF.

{\sl Note added.}--After the submission of this paper, we became aware of
a work~\cite{manna}, which starts from the same motivation as ours.

\vfil\eject
\end{document}